\documentclass{emulateapj}
\usepackage{color, hyperref, epsfig, natbib}
\usepackage{apjfonts}

\newcommand{\msun}{\ensuremath{M_{\odot}}}

\newcommand{\ha}{H\ensuremath{\alpha}}
\newcommand{\hb}{H\ensuremath{\beta}}
\newcommand{\oiii}{[O\,{\footnotesize III}]}
\newcommand{\oiv}{[O\,{\footnotesize IV}]}
\newcommand{\loiii}{$L_{\rm \oiii}$}
\newcommand{\lbol}{$L_{\rm bol}$}
\newcommand{\nii}{[N\,{\footnotesize II}]}

\newcommand{\kms}{\ensuremath{\mathrm{km~s^{-1}}}}

\newcommand{\sersic}{S\'{e}rsic}
\newcommand{\wise}{\emph{WISE}}
\def\kpc{\, h^{-1}{\rm {kpc}}}
\def\Mpc{\, h^{-1}{\rm {Mpc}}}

\begin{document}

\title{Differences in Halo-Scale Environments between Type~1 and Type~2 AGNs at Low Redshift}

\author{Ning~Jiang\altaffilmark{1}, Huiyuan~Wang\altaffilmark{1},
Houjun~Mo\altaffilmark{2,3}, Xiao-Bo~Dong\altaffilmark{4},
Tinggui~Wang\altaffilmark{1}, Hongyan~Zhou\altaffilmark{5,1} }
\altaffiltext{1}{Key laboratory for Research in Galaxies and Cosmology,
Department of Astronomy, The University of Science and Technology of China,
Chinese Academy of Sciences, Hefei, Anhui 230026, China; ~jnac@ustc.edu.cn,
whywang@mail.ustc.edu.cn}
\altaffiltext{2}{Department of Astronomy, University of Massachusetts,
Amherst MA 01003-9305, USA; ~hjmo@astro.umass.edu}
\altaffiltext{3}{Physics Department and Center for Astrophysics, Tsinghua University,
Beijing 10084, China}
\altaffiltext{4}{Yunnan Observatories, Chinese Academy of Sciences,
Kunming, Yunnan 650011, China; Key Laboratory for the Structure and Evolution of
Celestial Objects, Chinese Academy of Sciences, Kunming, Yunnan 650011, China;
~xbdong@ynao.ac.cn}
\altaffiltext{5}{Polar Research Institute of China, 451 Jinqiao Road, Shanghai, 200136, China}

\begin{abstract}
Using low-redshift ($z<0.09$) samples of AGNs, normal galaxies and
groups of galaxies selected from the Sloan Digital Sky Survey (SDSS),
we study the environments of type~1 and type~2 AGNs both on small and large scales.
Comparisons are made for galaxy samples matched in redshift, $r$-band luminosity,
\oiii\ luminosity, and also the position in groups (central or satellite).
We find that type~2 AGNs and normal galaxies reside in similar environments.
Type~1 and type~2 AGNs have similar clustering properties on
large scales ($\gtrsim1~\Mpc$), but at scales smaller than $100\kpc$,
type~2s have significant more neighbors than type~1s
($3.09\pm0.69$ times more for central AGNs at $\lesssim30\kpc$).
These results suggest that type~1 and type~2 AGNs
are hosted by halos of similar masses, as is also seen directly from the mass
distributions of their host groups ($\sim10^{12}h^{-1}\,\msun$ for centrals and
$\sim10^{13}h^{-1}\,\msun$ for satellites). Type~2s have significantly more satellites
around them, and the distribution of their satellites is also more centrally concentrated.
The host galaxies of both types of AGNs have similar optical properties, but
their infrared colors are significantly different. Our results suggest that
the simple unified model based solely on torus orientation is not sufficient,
but that  galaxy interactions in dark matter halos must have played
an important role in the formation of the dust structure that obscures AGNs.

\end{abstract}

\keywords{ galaxies: active --- galaxies: general --- galaxies: halos --- galaxies: interactions}

\section{Introduction}

It is widely believed that almost all massive galaxies have supermassive black holes (SMBHs) in their centers, and that
active galactic nuclei (AGNs) are SMBHs actively accreting surrounding materials.
Observationally, AGNs are classified into two populations, type~1 and type~2, depending on
whether broad emission lines appear in their optical spectra.
The most popular model hypothesizes 
that the two types are intrinsically the same, and the observed differences
between the two are attributed solely to orientation effects (Antonucci 1993).
In such a unified model, a type~1 AGN is assumed to be observed with a direct view of its
nucleus,  while the accretion disk and the broad-line
region (BLR) of a type~2 AGN are blocked by an optically thick obscuring structure, 
called 'torus'. However, evidence has emerged that 
the simple unified model may not be able to explain all of the
observational facts. Some revisions of the model, 
for example including the evolution of torus and the BLR (e.g., Laor 2003;
Elitzur \& Ho 2009; Gu et al. 2013; Elitzur et al. 2014),
are needed (see the review by Netzer 2015 and references therein).

If type 1 and type 2 AGNs differed only in their orientations, 
they would be expected to have similar circumgalactic environments on 
halo and larger scales. The large scale environments of AGNs are usually measured by 
the auto-correlation function of AGNs (e.g., Porciani et al. 2004; Wake et al. 2004; 
Croom et al. 2005; Myers et al. 2007; Shen et al. 2007; 
Ross et al. 2009; Eftekharzadeh et al. 2015; Chehade et al. 2016), 
the cross-correlation function between AGNs and galaxies (e.g., Li e al. 2006;
Coil et al. 2007; Hickox et al. 2009; Krumpe et al. 2010; Miyaji et al. 2011; 
Shen et al. 2013; Zhang et al. 2013; Shao et al. 2015), 
and their modifications (e.g., Dahari 1984; Schmitt 2001; Ellison et al. 2011; 
Kollatschny et al. 2012).  

To test the unified model, some early studies have examined the environments 
of AGNs, using relatively small AGN samples (typically a few tens to
$\sim100$ AGNs). These studies have found that type~2 AGNs tend to have
more close companions within $\sim 100\kpc$ than type~1s (e.g. Laurikainen \& Salo 1995; 
Dultzin-Hacyan et al. 1999; Krongold et al. 2002; Koulouridis et al. 2006). 
Based on the cross correlation between a large sample 
of local AGNs and photometric galaxies, Strand et al. (2008) found a similar but weaker trend 
on larger scales , $\sim 2\Mpc$. In addition to the number of companions,  Villarroel \& Korn (2014) found 
that the color and activity are systematically different between the neighbors of type~1 and type~2 AGNs, 
and that the spiral fraction of the host galaxies depends on the environment of type~1 and type~2 AGNs in 
different ways. All these are at odds with the expectation of the simple unified model.

More studies based on high redshift AGNs (usually called quasars) have also been carried out 
to examine the unified model. Unlike local AGNs, high-$z$ type~2 AGNs are selected according to their high 
absorption column densities or strong dust extinctions, inferred from both X-ray and infrared (IR) data. 
Early results found no obvious difference between type~1 (unobscured) and type~2 (obscured) AGNs
in their clustering,  giving support to the unified model (e.g. Ebrero et al. 2009; Coil et al. 2009; Gilli et al. 2009; 
Geach et al. 2013).  However, more recent studies suggested that the angular clustering amplitudes for the 
two types of AGNs are significantly different, with type~2s being more strongly clustered than 
type~1s  (e.g. Hickox et al. 2011; Elyiv et al. 2012; Donoso et al. 2014; DiPompeo et al. 2014; 
but see Allevato et al. 2011, 2014 for some different results), consistent with the results obtained 
for local  AGNs. Similar results were found by DiPompeo et al. (2015, 2016) using the cross
correlation between the cosmic microwave background (CMB) lensing map and the distributions
of the two AGN populations. All these studies suggest that type~2 AGNs tend to live 
in more massive halos. However the difference in the inferred halo masses vary greatly 
from study to study.

It should be emphasized, however,  that many early investigations have already revealed that 
AGN clustering depends on various properties, such as the luminosity (\lbol, e.g.,
Serber et al. 2006; Strand et al. 2008), redshift (e.g., Croom et al. 2005;
Zhang et al. 2013; Eftekharzadeh et al. 2015) and various other attributes of 
the host galaxies (e.g. Li et al. 2006; Coil et al. 2009; Hickox et al 2009; 
Mandelbaum et al. 2009; Mendez et al. 2016; see also \S\ref{sec_cont}). 
If the samples of the type~1 and type~2 AGNs used in the clustering analyses have 
different distributions in these properties, the results obtained may be 
biased and, therefore, are difficult to be used to constrain theoretical models, 
such as the unified model.

In this paper, we analyze the environmental dependence of AGNs using a
large, well-designed type~1 and type~2 AGN samples at low-redshift
selected from the Sloan Digital Sky Survey (SDSS). These AGN samples have
reliable measurements of AGN parameters and other information
about the host galaxies. With all these, we can build well-defined and
well-controlled samples for our analysis. Our primary goal is to compare the
environments of the two types of AGNs on both small and large scales
using cross-correlations between AGNs and reference galaxies.
Compared with previous studies of quasar clustering, our low-redshift
samples are much more suited to study environments on relatively small scales,
because here galaxies that are used to trace the environments of
AGNs  can be observed to faint luminosities.
We also use groups of galaxies to study the small scale environments
of AGNs, double-checking the results obtained from the cross-correlation analysis.
As we will show later, type~1 and type~2 AGNs reside in halos with similar mass 
distributions but have different number of galaxy companions within halos, which is inconsistent with
the assumption that type 1 and type 2 AGNs differ only in orientation,
and suggests a difference in physical mechanisms related
to triggering, fueling and obscuration between the two types of  AGNs.
To better understand the origin of this difference,
we also analyze the environments of matched normal (inactive) galaxies.

The paper is organized as follows.
In Section 2 we introduce the data and samples used in this paper.
In Section 3 we present our results concerning AGN environments, using
cross-correlation between AGNs and normal galaxies.
We analyze the properties of the host halos and host galaxies
of different types of AGNs in Section 4. Finally, we summarize our results
and discuss their implications in Section 5.

\section{Data and Samples}
\label{sec_samp}

The data used in this paper are primarily obtained from SDSS, using results
previously obtained both by ourselves and others.
To investigate the environments and hosts of AGNs,
we need to construct samples not only for AGNs (both type~1 and type~2),
but also for normal galaxies and for groups of galaxies.

\subsection{The Parent AGN Samples}
\label{ssec_AGN}

The parent AGN sample is selected from SDSS DR4 (Adelman-McCarthy et al. 2006).
The AGN classification is based on the traditional definition
that relies on the presence or absence of broad emission lines
in the AGN optical spectra. In Dong et al. (2012), a set of elaborate,
automated selection procedures was developed to conduct AGN--galaxy
spectral decomposition and continuum/emission-line fitting.
Dong et al. started with 451,000 spectra classified by the SDSS
spectroscopic pipeline as "galaxy" or "QSO" at $z<0.35$
(to ensure that \ha\ emission line is in the spectrum) in the SDSS DR4.
Since \ha\ emission is in general the strongest
broad line in the optical spectra of AGNs,
Seyfert 1s are identified according to the presence of broad \ha.
The criteria of the reliable detection of broad \ha\ are set quantitatively
according to the prominence of broad-\ha\ flux as well as
the statistical significance (see Dong et al 2012 for the detail).
This procedure yields a total of 8,862 broad-line AGNs at $z<0.35$
with secure detections of the broad \ha\ line. These AGNs
form the parent sample of type~1 AGNs adopted here.

The type~2 AGN sample is the one built by Dong et al. (2010).
It comprises 27,306 Seyfert~2 galaxies in the SDSS DR4
selected in the BPT line-ratio diagram (Baldwin, Phillips \& Terlevich 1981)
using the demarcation lines of Kauffmann et al. (2003).
All the narrow emission lines, \hb , \oiii~$\lambda 5007$, \ha\ , and \nii~$\lambda6583$
are required to be detected at $>5~\sigma$ significance.
In order to measure the emission lines reliably, the following procedure
is applied. First the starlight continuum, obtained from stellar templates
broadened and shifted to match the stellar velocity dispersion of the galaxy
in question, is subtracted to obtain a clean emission-line spectrum.
Note that the stellar absorption features are
thus subtracted, ensuring the reliable measurements of weak emission lines.
Each emission line is fitted incrementally with as many Gaussians as
statistically justified; basically, 2 Gaussians are used to model every doublet
line of \oiii~$\lambda\lambda4959, 5007$ and 1 Gaussian is
sufficient to model the others.  We refer the reader to
Dong et al. (2010; their Section~2.2) for the details of
the spectral fitting, sample construction, and measurements
of the AGN properties. The broad-line objects in the type~1 sample of
Dong et al (2012) are eliminated
even if they are selected according to the above criteria.

\subsection{Galaxies and Groups of Galaxies}
\label{ssec_groups}

The galaxy sample used here is the same as that used in
Yang et al. (2007, hereafter Y07) in the construction of their
galaxy group catalog\footnote{http://gax.shao.ac.cn/data/Group.html}.
It is based primarily on the New York University
Value-Added Galaxy Catalogue (NYU-VAGC; Blanton et al. 2005)
of the SDSS DR7 (Abazajian et al. 2009). Y07 selected all galaxies in the main
galaxy sample, with $r$-band Petrosian magnitudes $r\le17.77$ after correcting
for Galactic extinction (Schlegel et al. 1998), with redshifts in the range
$0.01\leqslant z \leqslant 0.2$, and with redshift completeness $C > 0.7$.
Among these galaxies, 599,301 have redshifts from the SDSS,
3,269 have redshifts taken from other redshift surveys, and
36,789 that lack redshifts due to fiber collisions
have assigned redshifts of their nearest neighbors.
It should be noted that, although this fiber collision correction
works well in roughly 60\% of all cases, the remaining 40\%
can be very different from their corresponding true values
(Zehavi et al. 2002). In this paper, the sample of galaxies with redshift
obtained from SDSS and other redshift surveys, as well as from
fiber collision corrections is referred to as sample A,
and the one including only measured redshifts (from both SDSS
and other sources) is referred to as sample B.

From these galaxy samples, Y07 constructed galaxy group catalogs,
using their halo-based group finder (Yang et al. 2005) and
taking into account various observational selection effects.
Each group is assigned a halo mass based on the ranking of its
characteristic luminosity/stellar mass. Following the original
definition, the brightest galaxy in a group is referred to as
the central galaxy while all others as satellites. Unless specified
otherwise, our results are presented for sample A and the group
catalog constructed from it.

\begin{figure}
\centering
{
\includegraphics[width=0.45\textwidth]{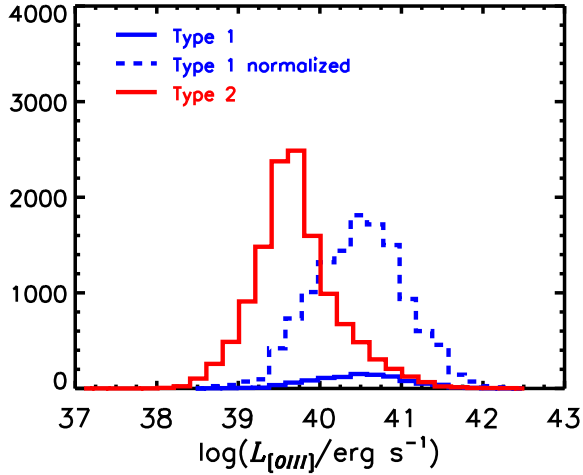}%
}
\caption{
The \loiii\ distributions of parent type~1 (blue) and type~2(red) samples at $z<0.09$.
Dashed blue line is the distribution for type~1 AGNs that is scaled to have the same number as type~2s.
}
\label{oiiidis}
\end{figure}

\subsection{Control Samples of AGNs and Normal Galaxies}
\label{sec_cont}

As noted above, each object in our parent type~1 AGN sample is identified either as
a "QSO" or as a "galaxy" in the SDSS pipeline. Because of the redshift limit
of our galaxy sample, many QSOs given by the SDSS pipeline do not
lie in the volume within which galaxy groups are identified. In fact,
the fraction of type~1 AGNs that are identified as galaxies by the SDSS pipeline
is close to unity at $z<0.09$ and decreases gradually with redshift. In our
analyses we will consider objects at $z\le 0.09$ so that galaxies with $r$-band
absolute magnitudes $M_r\le -19.5$ are complete. In this redshift range,
we have 1035 type~1 AGNs, about 94\% of the objects selected into
the parent sample.

As described in \S\ref{ssec_AGN}, the selection methods of
type~1 and type~2 AGNs are different and thus the comparison as a whole should not be fair.
Within the same redshift range ($z<0.09$), there are 12,621 type~2 AGNs,
about 12 times as many as the type~1s. To carry out a fair comparison,
a control sample of type~2 AGNs is constructed to match the properties
of the type~1 sample.
We choose to control four parameters: \oiii\ luminosity (\loiii),
$r$-band absolute magnitude ($M_r$), redshift ($z$) and
central/satellite classification. The reasons for these are the following.
\begin{itemize}
{\item {\it \loiii}: Some previous studies have shown that AGN clustering properties
depend on their \lbol\ (e.g., Li et al. 2006;
Serber et al. 2006; Strand et al. 2008). On the other hand,
it has been proposed that the covering
factor of the torus may be directly related to \lbol\, as in the so called
receding and approaching torus models (e.g.,
Lawrence 1991; Laor 2003; see also Netzer 2015 as a review).
So a control in \lbol\ is needed to deal with the effects of any difference
in \lbol\ between the type 1 and type 2 samples due to different torus
covering factors. \loiii\ is commonly
adopted as a good indicator of the \lbol\ of an AGN because it is
believed to originate from the narrow line region and to be only
weakly affected by the viewing angle relative to the torus
(e.g., Heckman et al. 2004; Lamastra et al. 2009).
As shown in  Figure~\ref{oiiidis}, the median of
\loiii\ for type~1 AGNs in our sample is higher than that of type~2s by $\sim 0.8$~dex.
To reduce any potential dependence on nuclear luminosity,
we first match type~1 and type~2 samples in their \loiii\ distributions. }
{\item {\it $M_r$}:
It is well known that the clustering amplitude of galaxies depends significantly
 on  galaxy luminosity, in that luminous galaxies tend to reside in more massive dark matter
halos than fainter galaxies (e.g., White \& Rees 1978;
Yang et al. 2003; Vale \& Ostriker 2004).
Several recent studies have indeed suggested that AGN clustering,
including the difference in clustering between obscured and unobscured
AGNs,  is simply determined by the luminosities of their host galaxies
(e.g., Mendez et al. 2016).
To avoid this effect, AGNs in our type 1 and type 2
samples are paired, so that the difference in the absolute magnitude
between the two galaxies in each pair $\vert\Delta M_r\vert <0.1$.}
{\item {\it z}: Redshift dependence of AGN clustering has been found
in a number of previous studies
(e.g., Croom et al. 2005; Zhang et al. 2013; Eftekharzadeh et al. 2015;
Chehade et al. 2016). To minimize possible such dependence, the two
samples are also paired in redshift, so that the redshift difference
between the two AGN in a pair  $\Delta z<0.01$.}
{\item {\it Central/satellite classification}:
This separation itself represents a characterization of the halo-scale environment
and it has been known that central and satellite galaxies
may evolve in different ways (e.g., Dressler 1980; Hashimoto et al. 1998).
Among the type~1 AGNs in our working sample, about 79.4\% are central galaxies according
to the group catalog.  The central fraction of type~2 AGNs is 79.1\%,
almost identical to that of type~1s. In matching type~1s and type~2s,
centrals are only matched with centrals, and satellites only with satellites.
 }
\end{itemize}

\begin{figure*}
\centering
{
\includegraphics[width=0.96\textwidth]{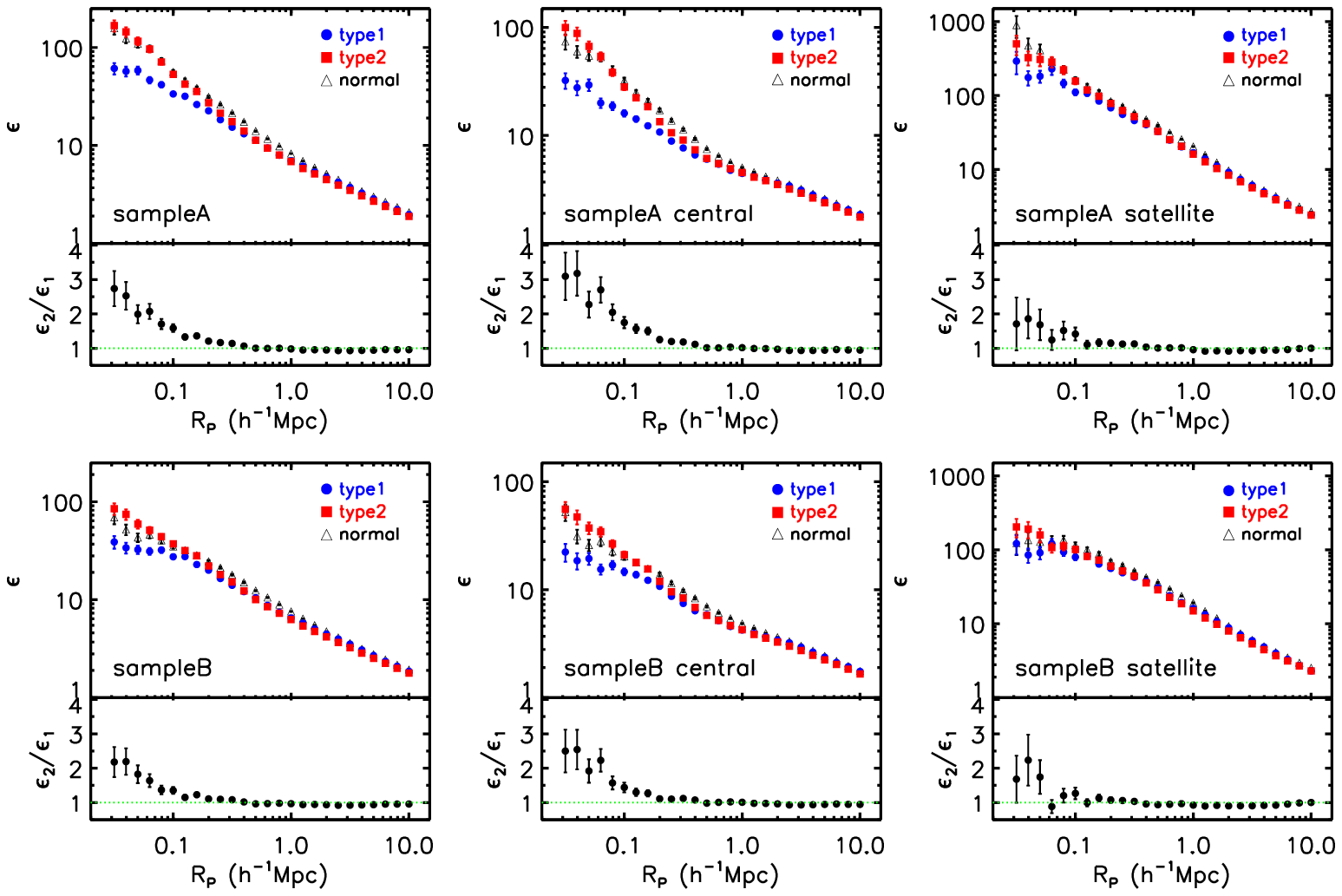}
}
\caption{In each panel, the top sub-panel shows the cross-correlations of control type~1
AGNs (red square), type~2 AGNs (blue circle) and normal galaxies (black triangle) with
reference galaxies, and the bottom sub-panel shows the ratio of cross correlations
between type~2 and 1 AGNs (black circle).
The errors are given by bootstrap method as described in \S\ref{sec_cc}.
The left panels show the results for all control AGNs
and normal galaxies, the middle panels for central galaxies and right panels for satellites.
The reference galaxies are constructed from sample A for the top three panels and
sample B for the bottom panels. Please see Section \ref{sec_samp} for the definitions
of these samples.
}
\label{cc}
\end{figure*}

In practice, for each type~1 AGN in our working sample, we select
five type~2s with $|\Delta M_r|<0.1$ and $|\Delta z|<0.01$,
which are in the same category (central or satellite) as,
and have the  \loiii\ closest to, the type~1 AGN in question.
The measurements of \loiii\ are drawn directly from
Dong et al. (2010, 2012). We choose five closest matches,
instead of one, to reduce the statistical uncertainties.
This is possible because we have many more type~2s than type~1s.
The difference in \loiii\  between the matched pairs
has a median value of zero and a variance $\sigma \sim 0.05$~dex.
Finally we obtain type~1 and type~2 samples that have similar distributions
in \loiii, $M_r$, redshift and central/satellite fraction.

In addition to compare type~1 and type~2 AGNs, we also want to
investigate the difference between active and normal galaxies
so as to understand the environmental difference of AGNs and
how AGNs are triggered and fueled.
We will therefore also analyze the properties of a control sample of
normal galaxies. For each AGN, a normal galaxy of the same
category (central or satellite) is selected from the
SDSS galaxy sample that has $M_r$ closest to
that of the AGN in question and a redshift difference less
than 0.01. Note that we do not eliminate active galactic nuclei
from the pool of  "normal" galaxies in matching
an AGN with another galaxy.  Since AGNs are only a small fraction
of all galaxies, including or excluding them in the matching pool
does not make a difference. For each type~1 AGN in the
working sample, five matches of normal galaxies are selected,
while for each object in the control sample of type~2 AGNs,
only one match is made.  Our tests show that these two control
samples give almost identical results for all the quantities
examined in this paper. We therefore combine them into one
control sample for  normal galaxies, which is 10 times as large
as the type~1 sample and 2 times as large as the control
type~2 sample.

\begin{deluxetable*}{c|cccccc}
\tablecaption{The Ratio of the Cross-Correlation between Type 2 and Type 1 AGNs ($\epsilon_2/\epsilon_1$)}
\tablewidth{0pt}
\tablehead{
\colhead{Sample}   &
\colhead{$R_P=10^{-1.5}$}   &
\colhead{$R_P=0.1$}   &
\colhead{$R_P=10^{-0.5}$}   &
\colhead{$R_P=1$}  &
\colhead{$R_P=10^{0.5}$}   &
\colhead{$R_P=10$}    \\
(1) & (2) & (3) & (4) & (5) & (6) & (7) }
\startdata
$\rm sampleA (sA)$        & $2.82\pm0.89$ & $1.56\pm0.12$ & $1.13\pm0.04$ & $0.98\pm0.02$ & $0.94\pm0.02$ & $0.95\pm0.02$  \\
$\rm sA~central$    & $3.09\pm0.69$ & $1.75\pm0.17$ & $1.18\pm0.06$ & $1.02\pm0.03$ & $0.94\pm0.02$ & $0.95\pm0.02$  \\
$\rm sA~satellite$  & $1.71\pm0.70$ & $1.42\pm0.19$ & $1.13\pm0.06$ & $0.96\pm0.04$ & $0.93\pm0.03$ & $1.00\pm0.03$  \\
$\rm sampleB (sB)$        & $2.18\pm0.53$ & $1.33\pm0.09$ & $1.08\pm0.04$ & $0.97\pm0.02$ & $0.93\pm0.02$ & $0.95\pm0.02$ \\
$\rm sB~central$    & $2.50\pm0.63$ & $1.44\pm0.14$ & $1.12\pm0.06$ & $1.01\pm0.03$ & $0.93\pm0.02$ & $0.95\pm0.02$  \\
$\rm sB~satellite$  & $1.68\pm0.61$ & $1.27\pm0.16$ & $1.03\pm0.06$ & $0.93\pm0.04$ & $0.91\pm0.03$ & $1.00\pm0.03$  \\
$\rm sA~central^{1000}$    & $2.80\pm0.52$ & $1.61\pm0.13$ &  $1.16\pm0.04$   & $1.00\pm0.01$ &  $0.93\pm0.01$ & $0.95\pm0.01$
\enddata
\tablecomments{
(1): The galaxy sample used to calculate the cross-correlation and for central/satellite
division.  Sample A refer to all galaxies with redshift obtained from SDSS and other
redshift surveys, as well as from fiber collision corrections; sample B only include
galaxies with spectroscopic redshift.
The cross-correlation function $\epsilon$ is defined as formula~\ref{fepsilon}.
The last row is also for the sample A central AGNs, while adopting $v_c=1000$~\kms.
(2)-(7): the $\epsilon_2/\epsilon_1$ with $R_P=10^{-1.5}, 0.1, 10^{-0.5},
1, 10^{0.5}, 10~h^{-1}$~Mpc, respectively. Their errors are estimated from the
bootstrap method.
}
\label{tb_cc}
\end{deluxetable*}

\subsection{Reference and Random Galaxy Samples}

One of our goals is to quantify the environments in which different types
of AGNs reside. Here we use galaxies as tracers of the environments.
To this end, we use a reference sample of 170,095 galaxies
with $M_r \leq -19.5$ at $z\leq 0.09$ in sample A.
The magnitude cut is chosen to ensure that the reference sample
is complete in the redshift range adopted for our control samples.

To account for the effects due to the irregular survey geometry,
we also generate a random sample which is 200 times as large
as the reference galaxy sample, with a total of 34,019,000 objects.
The redshifts and magnitudes of these random galaxies are exactly
the same as those in the reference sample, but with their coordinates
(right ascension, declination) randomly selected from a
uniform distribution in the sky. We determine whether or not
a random galaxy is in the SDSS footprint, using the {\tt IDL}
program `is\_in\_windows.pro' in {\it idlutils}. The geometry of the
footprint is described by a set of polygons, and the areas around
bright stars and sectors with ${\rm fgotmain}<0.7$
\footnote{see http://sdss.physics.nyu.edu/vagc/\#geometry}
are excised.

\section{Environments on Small and Large Scales}

\subsection{Cross-Correlation between AGNs and Galaxies} \label{sec_cc}

The cross-correlation between AGNs and galaxies has been used to analyze the
environments of AGNs (e.g., Li et al. 2006, 2008; Hickox et al. 2009).
Compared to the auto-correlation of AGNs, the cross-correlation
is statistically more robust, because the large number of reference
galaxies. More importantly,  environments on small scales (e.g., <100~kpc)
can only be studied by such cross-correlation analysis, because
AGN pairs of such small separations are rare.

We define the cross-correlation function $\epsilon$ as,
\begin{equation}
\label{fepsilon}
\epsilon(R_p)=\frac{N_{\rm R}}{N_{\rm G}}
\frac{DG(r_p<R_p,c\Delta z<v_c)} {DR(r_p<R_p,c\Delta z<v_c)}\,,
\end{equation}
where $\rm DG$ and $\rm DR$ are, respectively, the pair counts between
AGNs  and the reference (tracer) galaxies, and between AGNs and random galaxies,
with projected separation $r_p<R_p$ and redshift difference $|c\Delta z|<v_c$;
$N_{\rm G}$ and $N_{\rm R}$ are the total number of galaxies in the reference
and random samples, respectively. Thus, if AGNs were randomly distributed
with respect to galaxies, then $\epsilon=1$.
We estimate the statistical errors in the cross-correlation measurements
using the bootstrap method. To this end, we generate $N=1,000$ bootstrap
AGN samples, each of which consists of AGNs randomly picked from
original sample allowing  multiple selections of individual objects.
The pair counts, DG and  DR, are estimated for each of the bootstrap
sample, and their errors are given by the standard deviation of the
measurements among all the bootstrap samples. It is interesting
to note that the bootstrap error is almost identical to the Poisson error
on small scales, because the total number of  galaxies around each AGN
is very low, and count of neighbors for individual AGNs is typically
either 1 or 0.
In our analyses, we choose $v_c=500~\kms$, motivated by the fact that it is about
several times the virial velocity of a typical AGN host dark matter halo,
which has a mass $\sim10^{12} \msun$ (e.g., Padmanabhan et al. 2009;
Ross et al. 2009; Shen et al. 2013; see also our results below).
Our tests show that choosing an alternative value of $v_c=1000~\kms$
does not change our results significantly
(see the last row of Table~\ref{tb_cc}).

The cross-correlation results for type~1 and type~2 AGNs, together with their ratios
($\epsilon_2/\epsilon_1$), are presented in the top left panel of Figure~\ref{cc}.
We also list the ratios and their bootstrap errors at a number of typical
radii in Table~\ref{tb_cc}.
On large scales, the type~1 and control type~2 samples exhibit very similar clustering,
implying that the two types of AGNs, on average, reside in halos of similar masses (see
\S\ref{ssec_halos}). On small scales, however, type~2 AGNs
have much stronger clustering with galaxies than type~1s.
At projected separation of $10^{-1.5}~\Mpc~(31.6~\kpc)$,
the average number of companions around type~2 AGNs
is about $2.82\pm0.89$ times that around type~1s. The $\epsilon_2/\epsilon_1$
decreases with increasing $R_p$, reaching a constant value of about one
at $R_p>100\,\kpc$. This suggests that the spatial distribution of
galaxies around type~2 AGNs is more concentrated than
that around type~1s only on small scales, typically within
the virial radii of the host halos of AGNs.

We also perform the same analyses for the control sample of
normal galaxies (see also Figure~\ref{cc}). Their behavior
looks similar to that the type~2
sample on both small and large scales, in good agreement
with results obtained previously (see e.g. Li et al. 2008).
This similarity in the clustering between type 2 AGNs and normal
galaxies has been used to argue that the environments
of AGNs are not very different from that of normal galaxies.
However, our results show that this is true only for
type~2 AGNs, but not for type~1s.

\subsection{Centrals versus Satellites}

According to current galaxy formation model (see e.g. Mo et al. 2010), central
galaxies and satellite galaxies may have experienced different evolutionary
processes. It is therefore interesting to analyze the central and satellite
populations separately. In the middle and right panels of Figure~\ref{cc}, we
show the cross correlations of central and satellite AGNs
with the reference galaxies, respectively.  For central galaxies,  we again see that
type~2 AGNs are more strongly clustered on small scales than type~1s, and that
the two types have similar clustering amplitudes on large scales. However,
the clustering difference on small scales is now more prominent than the
total (central plus satellite) population. The ratio of the cross correlation function
between type~2s and type~1s becomes larger than three at small project
separations, and the signal extends to separations $\sim200\kpc$.
Here again the normal central galaxies have similar clustering amplitude
as type~2 centrals on both large and small scales. In contrast, for satellites
the difference between type~1 and type~2 is small and, indeed,
insignificant given the error bars.  Overall, satellites are more strongly clustered
than centrals on both large and small scales, as is expected from the fact that
they reside preferentially in more massive halos. These results demonstrate
clearly that the difference in environment between type~1 and type~2 AGNs
is mainly for the central population. Note that about 80\% of all the AGNs
in our samples are centrals.  In what follows, we will focus on the central
population.

\begin{figure*}
\centering
{
\includegraphics[width=0.95\textwidth]{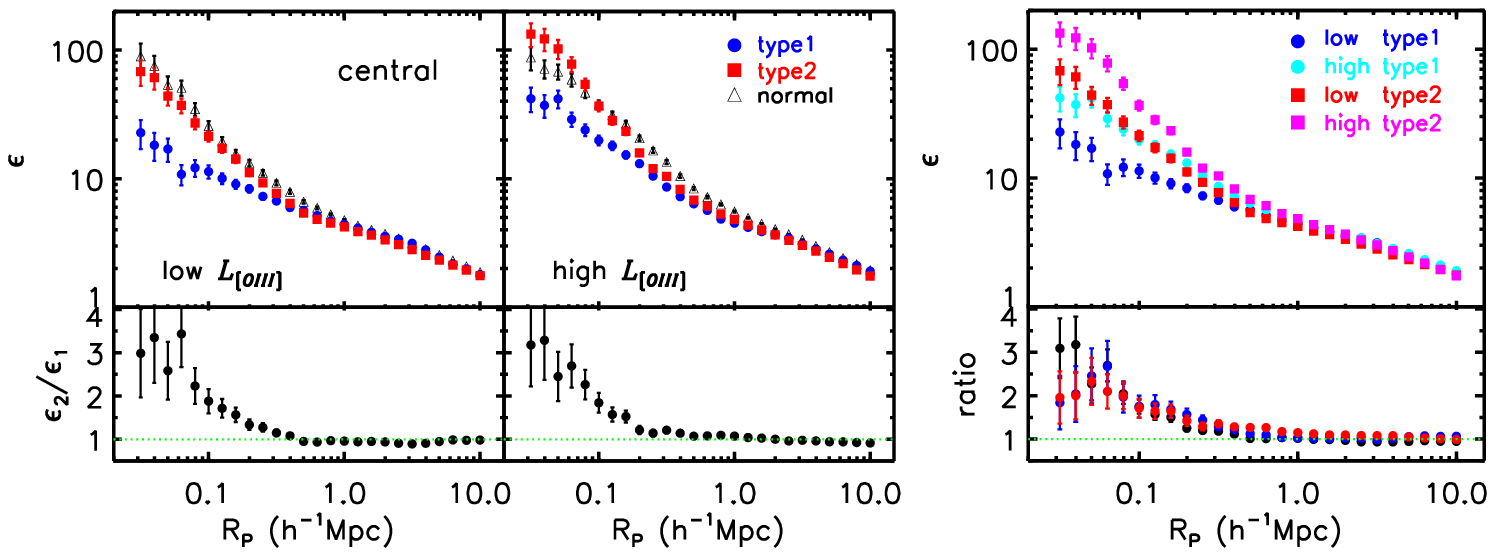}%
}
\caption{
Similar to Figure \ref{cc}. The left (middle) panel shows the results for central AGNs
of low (high) \loiii\ AGNs and their corresponding control normal galaxies.
The right panel presents the results for total (low and high \loiii) central AGNs.
In the bottom sub-panel of right panel, the blue circles represent the ratio
between high \loiii\ and low \loiii\ type~1 AGNs; and the red circles are the ratio
between high \loiii\ and low \loiii\ type~2 AGNs. We also plot the ratio between type~2
and type~1 AGNs as the black circles for comparison.
}
\label{ccoiii}
\end{figure*}

\begin{figure*}
\centering
{
\includegraphics[width=0.7\textwidth]{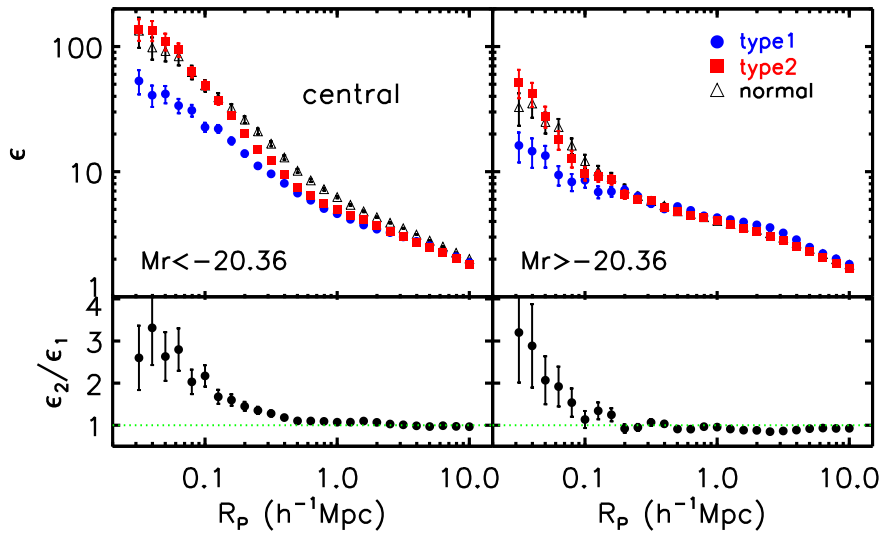}%
}
\caption{
Similar to Figure \ref{cc}. The left and right panels show the results for high- and
low-luminosity samples of central AGNs and their corresponding control normal
galaxies, respectively.
}
\label{ccmr}
\end{figure*}

\subsection{Dependence on  \loiii}\label{sec_loiii}

Environmental dependence of AGN luminosities has been investigated before
(e.g. Serber et al. 2006; Strand et al. 2008). It is found that AGNs of
higher luminosities tend to reside in denser environments.
Here we investigate whether the difference in clustering properties between type~1 and type~2
AGNs also depends on AGN activities (as indicated by their \loiii).
To do this we divide each of the type~1 and 2 samples (only central galaxies)
into two equal-sized subsamples according to the value of \loiii.
The median \loiii\ of the high-\loiii\ subsample is $\sim0.7$~dex higher
than the low-\loiii\ subsample.
The cross correlation results of these subsamples are presented in Figure~\ref{ccoiii}.
The difference between the two types of AGNs on small scales is observed
for both high- and low-\loiii\ samples, with type~2 AGNs having
a higher cross correlation amplitude than type~1s. On large scales,
the clustering amplitudes for the two types are quite similar, again
suggesting similar host halo masses for both types.
In the right panel, we plot the results for the high and low \loiii\ subsamples together.
For a given type, the clustering amplitude increases significantly with  \loiii\
at $R_p< 200\kpc$ but no such increase is seen at larger separations.
Such dependence is consistent with the results found in Strand et al. (2008).

\subsection{High versus Low Galaxy Luminosity}

In addition to the nuclear luminosity, the clustering amplitude of galaxies
are also known to depend on galaxy luminosity, with intrinsically
brighter galaxies having stronger clustering.
It is therefore interesting to check whether or not the clustering difference
between type~1 and type~2 AGNs also depends on the luminosities of host galaxies.
To do this, we split each of our central AGN samples into two subsamples
of equal size (in number)  according to the $r$-band absolute magnitude
($M_r$) of the host galaxy. The cross correlation results for these subsamples
are presented in Figure~\ref{ccmr}. Clearly, the difference on small scales
($<100\kpc$) and the similarity on large scales for the two types of AGNs
are seen for both high and low luminosity subsamples. However,
there are some noticeable differences in the results
between the two subsamples. First, the clustering amplitude of the higher
luminosity subsample is higher, consistent with the
results of normal galaxies (e.g. Wang et al. 2007; Guo et al. 2010;
Zehavi et al. 2011). Second, the clustering difference is noticeable
only at $<100\kpc$ for the lower luminosity subsample,
but extends to $\sim300\kpc$ for the higher luminosity subsample.
This may be understood if environmental effects that separate
type~1 and type~2 AGNs operate on halo scales, because
brighter galaxies reside preferentially in more massive galaxies
which have larger virial radii. Finally, the cross correlations of normal
galaxies in both of the luminosity ranges follow those of the corresponding
type~2 subsamples.

\begin{figure}
\centering
{
\includegraphics[width=0.48\textwidth]{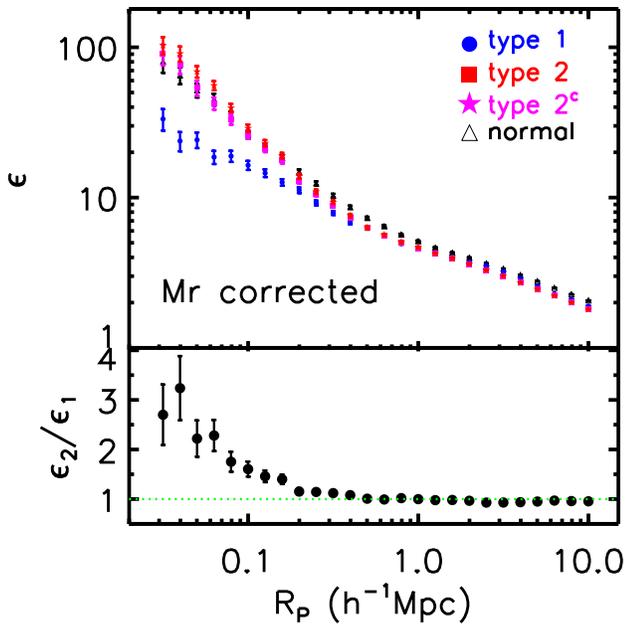}%
}
\caption{
Similar to Figure \ref{cc}. In order to estimate the AGN contamination, for each of the type~1 AGNs,
we remove the contribution of the AGN continuum to obtain a new $M_r$.
Based on this corrected $M_r$, we construct new control samples of type~2 AGNs
(type~2$^c$, magenta stars) and normal galaxies. For comparison, we also plot the
results for the old sample of the type~2 AGNs (red square). The bottom panel shows the
ratio between type~2$^c$ and type~1 AGNs. Only results for central galaxies are presented.
}
\label{ccc}
\end{figure}

\subsection{Testing the Impact of Selection Effects}
\label{ssec_testing}

Is it possible that the clustering difference between type~1 and 2 AGNs on
small scales is actually caused by some observational effects, rather than by
real environmental effects? Because of fiber collisions, a few percent of the
reference galaxies have no spectroscopic redshifts, and in Sample A
they are assigned the redshifts of their nearest neighbors. As mentioned above,
about 40\% of the assigned redshifts are not reliable (Zehavi et al. 2002).
In order to check whether or not such uncertainty is able to produce the
clustering difference between type~1 and type~2 AGNs, we have repeated
our analyses but using sample B, which only contains galaxies with spectroscopic
redshifts. The results are shown in the bottom panels of Figure~\ref{cc}.
As expected, the clustering amplitudes are reduced on small scales ($<100\kpc$)
due to the elimination of close pairs by fiber collisions, but
almost no change is seen on larger scales (see Li et al. 2006).
However, the difference between type~1 and type~2 AGNs remains
but is slightly reduced. The reduction is expected because type~2
AGNs, being more clustered with other galaxies on small scales,
are more strongly affected by fiber collisions in their close pair counts.

Since the clustering strength is expected to depend on galaxy luminosity,
the difference in the clustering may also be produced  if the
host galaxy luminosities of type~1 AGNs are systematically lower
than those of type~2s. As an attempt to control this effect,
we have matched their luminosity distributions in our control samples
(see \S\ref{sec_cont}).  However, the values of $M_r$ directly measured
also include the contributions from AGNs themselves, and such
contributions may be important for type~1 AGNs.  To quantify the extent of
AGN contamination in the observed luminosity,
we define a parameter $\theta=f_{\rm AGN}/f_{\rm total}$, where $f_{\rm AGN}$ and
$f_{\rm total}$ are the $r$-band flux of the AGN and the total flux, respectively.
The $r$-band flux from the AGN is obtained by convolving the AGN component
from our spectral decomposition (Dong et al. 2012) with the SDSS $r$-band filter
throughput curves and $K$-correcting it to $z=0.1$. We found that $\theta$
for type~1 AGNs has a median value of $0.06$ and a maximum of about $0.45$.
More than 90\% of all the type~1 AGNs in our sample
have $\theta<0.2$, consistent with
the results obtained previously (e.g., Reines \& Volonteri 2015), suggesting that
any bias introduced is only moderate. This is also consistent with the fact
that type~1 AGNs have very similar clustering properties on large scale to
the corresponding control samples of normal galaxies that are matched in $M_r$.
In order to test the effect more precisely,  we construct a new control type~2 AGN
sample and a normal galaxy sample, both matching the corrected $M_r$
distribution of the type~1 AGNs, with the AGN contributions to the
luminosities subtracted. The new cross correlation results for the
samples so matched, with matchings in other quantities the same as before,
are presented in Figure~\ref{ccc}. For comparison, the cross correlation
for the old control sample of type~2 AGNs is also plotted.
The results change little in the new matching,
and the clustering difference on small scale ($<100\kpc$) is almost the same
as before. We thus conclude that AGN contribution to the total
galaxy luminosity is not the reason for the observed clustering difference between
the two types of AGNs.

As shown above, the amplitude of AGN clustering increases with \loiii.
To take into account this effect, the control type~2 AGNs are matched
in \loiii with the type~1  sample. However, it may be possible that
the torus can obscure the inner part of the \oiii\ emission line regions
(e.g. Netzer et al. 2006; Zhang et al. 2008). This obscuration may be
more important for type~2 AGNs, and so the intrinsic \loiii\ of type~2 AGNs
may be systematically underestimated. If the extinction of \oiii\ is
sufficient enough, the clustering difference between the two types of AGNs
may be entirely due to dust extinction.  Unfortunately, the extinction is hard
to estimate. By comparing X-ray luminosity and mid-infrared
line \oiv\ luminosity with \loiii, some earlier studies have
suggested that the intrinsic \loiii\ of type~2 AGNs can on average be
two times  as high as the measured value (e.g., Netzer et al. 2006; Kraemer et al. 2011).
However, if anisotropy in the X-ray emission is taken into account
(e.g. Liu et al. 2014), the dust extinction may be much smaller.
In Section \ref{sec_loiii}, we have divided AGNs into high and low \loiii\
subsamples. The median \loiii\ of the high-\loiii\ subsample is $\sim0.7$~dex higher than
the low-\loiii\ subsample. The luminosity difference between the two subsamples
is therefore much larger than the estimated extinction of \loiii\ for type~2 AGNs.
The ratio of the cross correlations between the two subsamples
is smaller than the ratio between type~1 and 2 AGNs (the right panel of
Figure \ref{ccoiii}). Thus, even if \loiii\  of each of the type~2 AGNs
is reduced by a factor of two by obscuration,  the effect is still far too small
to reproduce the difference between the two types of AGNs.
Moreover, we also divide all type~2 AGNs into a sequence of subsamples
with \loiii\ successively increased by a factor of 2 between adjacent
subsamples. The clustering difference between adjacent subsamples are much
smaller than that between the two types of AGNs. All these tests suggest that
the underestimation of the intrinsic \loiii\ for type~2s
cannot change our results significantly.

\begin{figure*}
\centering
{
\includegraphics[width=0.85\textwidth]{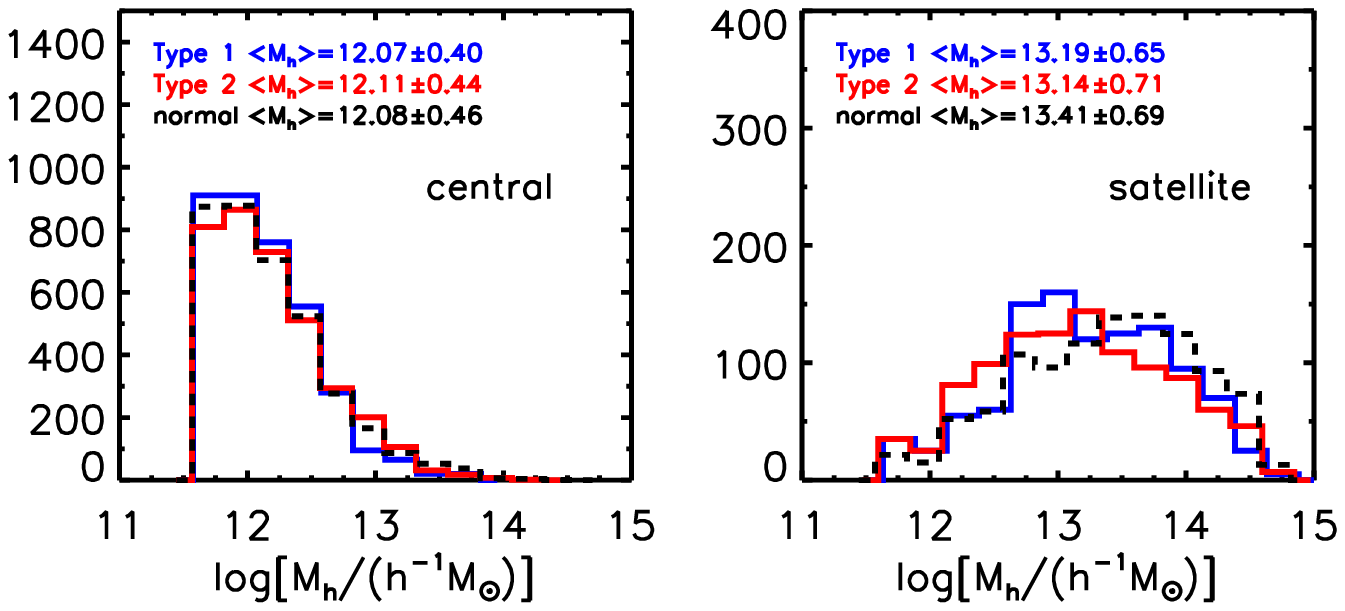}%
}
\caption{
Host halo mass distributions of type~1 (blue) and type~2 (red) AGNs.
Results for normal galaxies (black) are also plotted for comparison.
Left panel: central AGNs; Right panel: satellite AGNs.
The data indicated in the panels are the medians and
standard deviations of the distributions. For each type~1 AGN,
five type~2 AGNs and ten normal galaxies are matched (see Section 2.3).
To facilitate comparison, the distributions for type~1 AGNs and normal
galaxies are scaled by a factor of 5 and 0.5, respectively.
}
\label{AGNmh}
\end{figure*}

\section{Properties of Host Groups and Host Galaxies}
\label{sec_group}

The clustering difference indicates that the hosts of type~1 and type~2 AGNs
may have different properties. Here we examine the properties of their hosts directly.

\subsection{Properties of Host Halos}
\label{ssec_halos}

We use galaxy groups as given in Yang et al. (2007; see \S\ref{ssec_groups})
to represent dark matter halos. We first compare the mass distributions of halos
in which different types of AGNs reside (see Figure ~\ref{AGNmh}).
We see that type~1 and type~2 AGNs have very similar halo mass distributions,
consistent with the inference from their clustering properties on large scales.
For central AGNs, both distributions peak around $10^{12}h^{-1}\,\msun$,
in good agreement with results obtained previously for quasars
(e.g. Richardson et al. 2012; Shen et al. 2009, 2013) and from
galaxy groups (Pasquali et al. 2009). The halo masses of satellite AGNs are
on average one order of magnitude higher than central AGNs,
again consistent with the results of our clustering analyses.
Note that the halo masses of the groups are estimated by ranking the
total luminosity of member galaxies above a given luminosity, and
some small halos are not assigned halo masses (see Y07 for more
details). About $10\%$ of the central AGNs and $2\%$ of satellite AGNs in our
samples do not have assigned halo masses. These systems, with halo masses
all below $10^{11.6}h^{-1}\msun$ according to abundance matching,  are
not included  in the plot.

\begin{figure}
\centering
{
\includegraphics[width=0.48\textwidth]{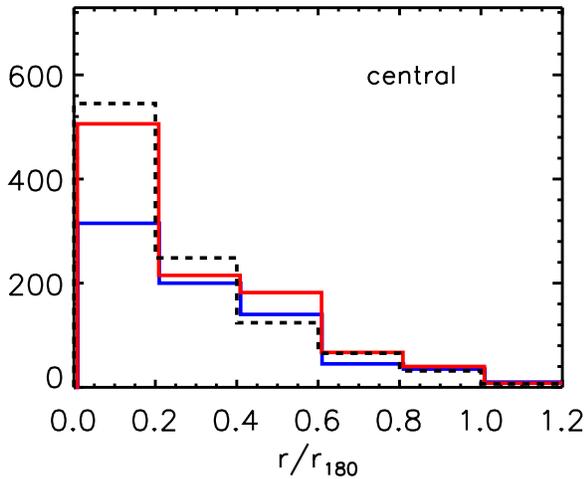}%
}
\caption{
The projected distance distribution of the closest satellites in the groups in which
type~1 or 2 AGNs are central galaxies. The distance is normalized by the virial
radius $r_{180}$. Similar to Fig. \ref{AGNmh}, the distributions for type~1 AGNs
and normal galaxies are scaled by a factor of 5 and 0.5, respectively.
}
\label{dclose}
\end{figure}

Next we check the number of satellite galaxies in the groups
where AGNs are hosted by the central galaxies. On average, there are
347/822=0.42 satellites in each of the type~1 AGN group and
2089/4110=0.51 in each of the type~2 AGN groups.
Here satellites with $M_r\leq-19.5$ and $M_r>-19.5$ are counted
separately. If only bright satellites of $M_r<-19.5$ are used,
the average numbers are 0.20 and 0.26 for type~1 and type~2,
respectively. We have also examined the properties of these
satellites, such as their $g-r$ color and \sersic\ index, but
no significant difference is found between halos hosting
type~1 and type~2 AGNs.

The slopes of the cross-correlations on small scales suggest that
the spatial distribution of galaxies around type~2 AGNs tend to be
more centrally concentrated, in comparison to that around type~1 AGNs.
To quantify this we measure the distribution of AGNs in terms of
the projected distance each of them has to the closest satellite. Note again
that almost all pairs ($\sim97\%$) have $c|\Delta z |\le 500\kms$.
The distributions for different types of AGNs and normal galaxies are presented
in Figure~\ref{dclose}. To reduce the dependence on halo mass,
the distance is normalized by the halo virial radius $r_{180}$
(see equation 5 in Y07). The distributions are not very different
at $r/r_{180}>0.2$, although type~2s appear to have systematically
more neighbors than type~1s up to $r/r_{180}\sim 1$.
However, at $r/r_{180}<0.2$, the number of pairs
for type~2 AGNs and normal galaxies are $\sim1.7$ times that
of type~1 AGNs. This clearly shows that type~2 AGNs
on average have higher number of close pairs than type~1s.

\begin{figure*}
\centering
{
\includegraphics[width=0.8\textwidth]{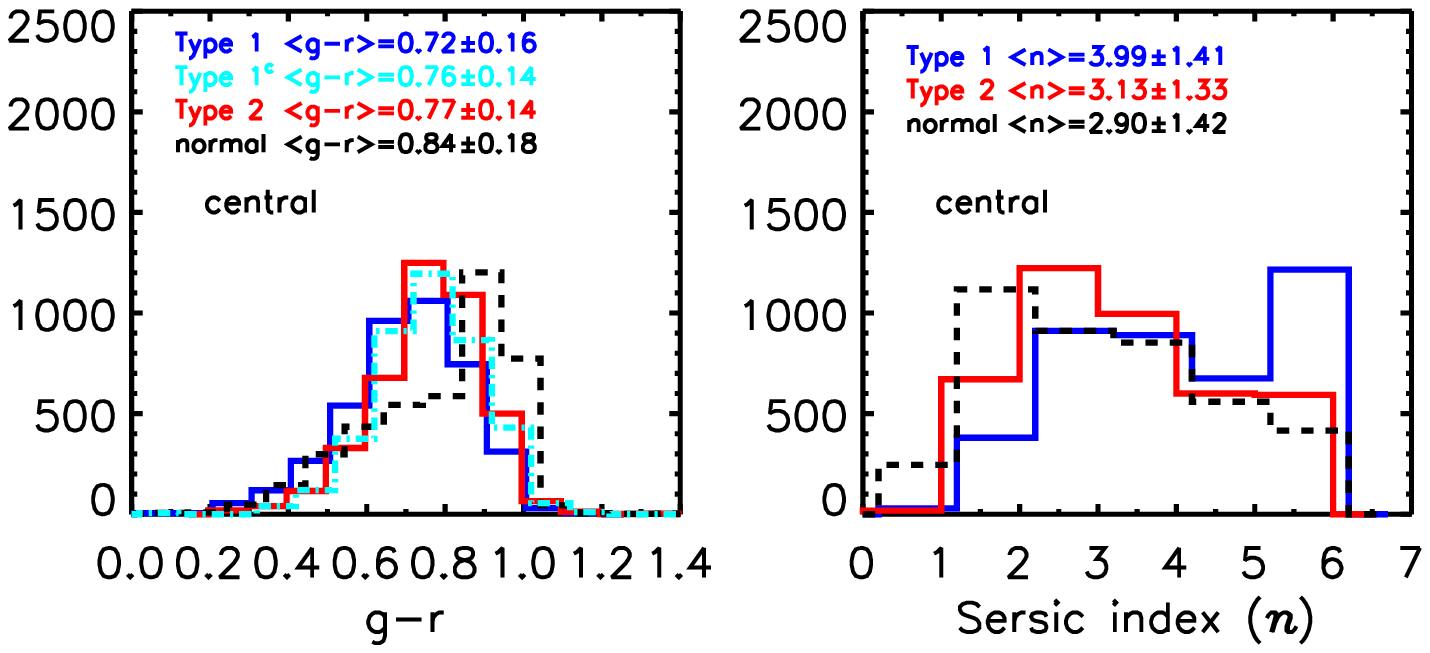}%
}
\caption{
The distributions of host galaxy properties for central type~1 (blue) and
type~2 (red) AGNs, and normal galaxies (black). Left: $g-r$ color; right: \sersic\ index.
We also show the host galaxy color distribution after AGN continuum is
subtracted for type~1 AGNs (cyan dot-dashed line).
The numbers indicated in the panels are medians and standard deviations
of the distributions. Similar to Fig. \ref{AGNmh}, the distributions for type~1 AGNs
and normal galaxies are scaled by a factor of 5 and 0.5, respectively.
}
\label{host}
\end{figure*}

\subsection{Properties of Host Galaxies}

The $(g-r)$ color distributions of the host galaxies of
the two types of AGNs as well as normal galaxies (only central galaxies)
are shown in Figure~\ref{host}. Type~1 AGNs on average are
slightly bluer than type~2s. Since AGNs are blue and
type~1 nuclei are more dominating in their hosts, the color
difference seen between type~1 and type~2 AGNs might be
produced by contaminations of the luminosities of the nuclei.
To test this, we have re-calculated the $(g-r)$ colors of type~1
AGNs by subtracting the contributions of the nuclei from the total
luminosities, using the method described in \S\ref{ssec_testing}.
We found that after this correction, the color difference between
type~1 and type~2 AGNs becomes negligible.
Both types of AGNs are bluer than normal galaxies. This is in
agreement with the fact that AGNs are found to reside preferentially
in the so-called ``green valley" galaxies, with colors intermediate between
star-forming blue cloud and the red sequence of galaxies
(e.g., Nandra et al. 2007; Salim et al. 2007).

For reference, we also show the \sersic\ index ($n$) distributions in the
right panel of Figure~\ref{host}. Type~1 AGNs show higher values of $n$,
meaning that they have more concentrated light distribution.
However, the high concentrations may be entirely due to the contributions
of the relatively bright nuclei. In order to eliminate these contributions,
careful image decompositions are needed.

\begin{figure*}
\centering
\includegraphics[width=0.6\textwidth]{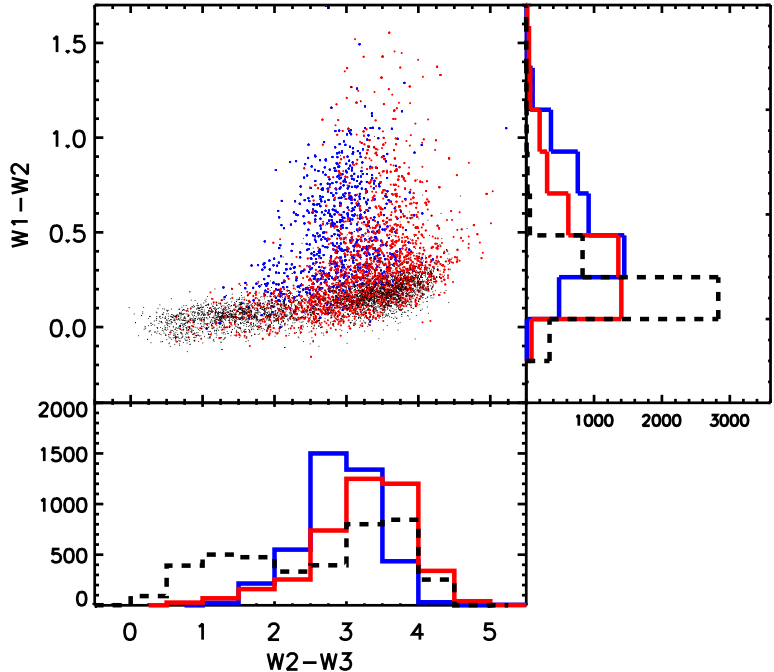}%
\caption{
The $WISE$ (W2-W3) versus (W1-W2) color diagram.
Blue: type~1 AGNs; red: type~2 AGNs; black: normal galaxies.
The loci of various types of objects seen here
are consistent with those in Figure~12 of Wright et al. (2010) and
Figure~3 of Alatalo et al. (2014).
Bottom: W2-W3 histogram; right: W1-W2 histogram.
Similar to Fig. \ref{AGNmh}, the distributions for type~1 AGNs and normal galaxies
are scaled by a factor of 5 and 0.5, respectively.
}
\label{wisediag}
\end{figure*}

The infrared color can also be used to study objects driven by
different physical processes. We have cross matched galaxies in
our working samples (central galaxies only) with \emph{Wide-field Infrared Survey Explorer}
(\wise , Wright et al. 2010) galaxies within a radius of 5\arcsec,
the spatial resolution of \wise\ in the near-infrared.  Almost all ($>99\%$) AGNs
and normal galaxies in our working samples are matched to \wise sources.
Figure~\ref{wisediag} shows the \wise\ (W2-W3) versus (W1-W2)
color-color diagram of the two types of AGNs
as well as the normal galaxies in the control sample,
where W1, W2 and W3 are the infrared magnitudes at 3.4, 4.6
and 12~$\mu$m, respectively.  Normal galaxies show bimodal
distribution in the (W2-W3) color, with the cloud
in the left dominated by early-type galaxies and
the cloud in the right by late-type galaxies. Their (W1-W2) color
distribution is rather narrow, peaked around 0.1. This suggests that
the warm dust heated by star-formation in late type galaxies emits
infrared photons primarily in the W3 band or bands of larger
wavelength. As a result, the  (W1-W2) color is dominated by star
light and is almost indistinguishable between early and late type galaxies.

In contrast, AGNs populate a large fraction in the region of
${\rm (W1-W2)}>0.5$. Note that ${\rm (W1-W2)}=0.5$ has been
suggested as a demarcation line between AGNs and normal
galaxies (e.g. Wright et al. 2010), on the basis that the photons
emitted by hot dust heated by AGN have higher energy than
that heated by stars. Our result is consistent with this demarcation
but shows that a significant number of AGNs, in particular type~2s,
have ${\rm (W1-W2)}<0.5$.  Furthermore, the result also shows that
type~1 AGNs are systematically bluer (and redder) than type~2 AGNs in the
(W2-W3) (and (W1-W2)).  This may be understood as a result
of a larger opening angle of the hot dust component that
can be seen for type~1 AGNs. It may also be possible that
the dust components in type~1 AGNs are systematically hotter
than those in type~2 AGNs.

\section{Summary and Discussion}

The study of the differences in environments between type~1 and
type~2 AGNs can play an important role, not only in testing the unified model,
but also in offering an avenue to explore the triggering/fueling
processes of AGNs and the connection between supermassive black holes
and their host galaxies.
In this paper we have re-visited this problem by using large uniform samples
selected from the SDSS, and by investigating the differences
between these two types of AGNs in their clustering, host halo and
host galaxy properties.

We find that, when type 1 and type 2 AGNs are matched in redshift,
$M_r$, and \loiii, the clustering strength of type~1 AGNs is almost
the same as that of type~2s on large scales, but much weaker
on scales smaller than about $100\kpc$. The clustering properties
of type 2 AGNs are similar to that of normal galaxies
with matched luminosities. Our results suggest
that dark halos hosting type~1 AGNs, on average, have
similar masses as those hosting type~2s, but that type~1s
have less satellites around them. The same conclusion is reached
by using galaxy groups to represent dark halos. In addition,
the distribution of the satellites around type~2 AGNs are more centrally
concentrated than those around type~1 AGNs.
We examine various selection effects to test the reliability of our results,
and find that none of the known effects can affect our results significantly.

The differences between type 1 and type 2 AGNs in their small scale clusterings
provide important information about these two populations of AGNs.
In the standard model, a torus-like, dusty structure is invoked
to unify the two populations, and the difference in the obscuration
produced by different inclination angles of the symmetric axis of the torus
relative to the observational line of sight is considered to be the only
reason for the discriminations of the two populations. In this case,
no difference is expected between type 1s and type 2s in their
environments,  clearly in conflict with what we find.
Our results are also different from those in some previous studies,
where it was found that the two types of AGNs reside in halos
of different masses but for high-luminosity quasars at higher redshift
(e.g., Allevato et al. 2014; DiPompeo et al. 2014),
and that there is no difference between AGNs and normal
galaxies in their environments (e.g., Ebrero et al. 2009; Coil et al. 2009).
It is important to note, however
high-luminosity quasars at high redshift may be different from
their local low luminosity counterparts, and one needs to keep this
in mind when comparing the low and high redshift objects.

The fact that type~1 and type~2 AGNs have different clustering properties only on small
scales suggests that galaxy interaction within dark halos may
play an important role in affecting the properties of AGNs.
In the unified model, the probability of a AGN to be observed as a
type~2 is proportional to the covering factor of the torus, and
so an AGN is more likely to be observed as a type~2 if
the torus have a larger covering factor. Thus, if the environmental
effects are to change the covering factor of the torus, the observed
difference in the small scale clustering between type~1s and type~2s
may be explained in the framework of the unified model.

The question is, of course, how the interactions on galactic ($100\kpc$)
scales can affect the gas/dust distribution on the torus
scales that is $10^5$ times smaller.
In the co-evolution scenario for SMBHs and their host galaxies
(see reviews by Kormendy \& Ho 2013 and Heckman \& Best 2014),
AGNs are triggered by interaction with nearby galaxies
(e.g., Dahari 1984; Sanders et al. 1988; Ellison et al 2011; Hong et al. 2015),
which can cause cold gas/dust to lose angular momentum and  flow into the
central region of the interacting galaxies. 
As shown by high-resolution hydrodynamic simulations
(e.g. Hopkins et al. 2012), when the amount of inflow gas is
sufficiently large, a lopsided and eccentric inner disk can form
and cause gas to move inwards to the central black hole, eventually
forming a torus. This scenario is supported by recent
observations (e.g., Shao et al. 2015), and consistent with our result
that type 2s tend to be more strongly correlated with other galaxies
so as to be more frequently affected by galaxy-galaxy interactions.

Alternatively, the interstellar medium (ISM) in the interacting galaxies
may be denser and dustier, and so optically thicker in dust
obscuration  than in normal galaxies. Thus, a fraction of
the type~2 AGNs may be obscured by galactic-scale dust distribution
rather than a torus, and the difference in clustering between
type~1 and type~2 AGNs may then be explained by the higher
number of interacting partners in the host halos of type~2 AGNs.
There are some indications for the presence of galactic-scale dust
obscuration. For example, Chen et al. (2015) found that the obscuration
at optical and X-ray bands in type~2 quasars is connected to
the far-IR-emitting dust clouds usually located far from the central
engine. In nearby low-luminosity AGNs, high resolution observations
have also revealed kpc-scale dusty filamentary structures that
are connected to dusty features close to the nucleus (Prieto et al. 2014).
Further evidence comes from observations that Compton-thick AGNs
are more likely hosted by galaxies with visible galactic dust lanes
(e.g. Goulding et al. 2012; Kocevski et al. 2015). Clearly, more
investigations about  the dust absorption properties of type~2 AGNs are
needed in order to distinguish between galactic-scale and torus obscuration.

If galaxy interaction indeed plays an important role in producing
different types of AGNs,  we may expect to observe  some interacting
signatures in the host galaxies.
Unfortunately, our inspection of the host galaxies
did not provide any reliable evidence for the difference between
type~1 and 2 AGNs, partly due to the contamination of AGN continuum in
type~1 objects. Clearly, high quality imaging data are needed to identify
signatures of galaxy interactions in these objects.
We should emphasize, however, that the galaxy interaction scenario
discussed here is different from the popular quasar evolutionary model,
in which violent mergers are assumed to be the trigger of AGNs.
Moreover, the AGNs and their hosts considered here have moderate
luminosities and masses, while quasar activities are probably associated
with more massive galaxies.
In that quasar case, a AGN may initially be heavily obscured by dust and
appear as a type 2 quasar. The AGN feedback may subsequently blow
away the surrounding gas and dust and evolve into an unobscured type 1
quasar. The final product of such an evolution is expected to be
a red and dead early-type galaxy (Sanders et al. 1988; Hopkins et al. 2006, 2008).
This is certainly not the kind of (the relatively weak) interactions we
are suggesting here for the low-$z$ AGNs. Furthermore.  the timescale of
galaxy mergers is typically Gyr (e.g., Boylan-Kolchin et al. 2008), much longer
than that of the AGN activities ($\lesssim0.01-0.1$~Gyr) we are concerned here.

Finally, we discuss another possibility to interpret our results.
It is well known that the inner structure of a dark matter halo is related to its
assembly history (see e.g. Wang et al. 2011): substructures
in a halo that assembled earlier tend to be destroyed by tidal stripping
or have fallen onto the central objects due to dynamical friction.
If, for example, type~1 AGNs are preferentially hosted by halos that
formed earlier, the amounts of sub halos, which themselves
host satellite galaxies and can interact with the central,
may be smaller in the host halos of type~1s than in those
of type~2s. This may also explain the difference in clustering between
the two types of AGNs. Recently, Lim et al. (2016) used the mass ratio
of the central galaxy to its host halo as an observable proxy of halo
assembly time and found it is
correlated with many properties of the galaxies it hosts.
Moreover, halo assembly history and substructure fraction are
found to depend on large scale structures (e.g. Gao et al. 2005; Wang et al. 2007),
an effect usually referred to as assembly bias. Thus, this scenario based on halo
assembly can be tested by studying the properties of the host halos
of AGNs in detail using galaxy groups, such as those given by
Yang et al. (2007).

\acknowledgments
We thank the anonymous referee for a thorough report and many constructive
comments that helped us improve the presentation of this work.
This work is supported by 973 program (2015CB857005, 2012CB821804), NSFC
(11522324,11421303),  the Strategic Priority Research Program "The
Emergence of Cosmological Structures" of the Chinese
Academy of Sciences, grant No. XDB09010400 and the Fundamental
Research Funds for the Central Universities. H.J.M.
would like to acknowledge the support of NSF AST-1517528.
The numerical calculations have been done on the supercomputing system in the
Supercomputing Center of University of Science and Technology of China.

\end{document}